\documentclass[journal]{IEEEtran}
\usepackage{cite, bm, geometry, parskip, psfrag, subfigure, url, stfloats, array, float, graphicx, epstopdf, color, amsmath, amssymb}
\newtheorem{thm}{Theorem}

\newtheorem{lem}{Lemma}
\newtheorem{prop}{Proposition}

\DeclareMathAlphabet{\eurm}{U}{eur}{m}{n}
\DeclareMathAlphabet{\mathbsf}{OT1}{cmss}{bx}{n}
\DeclareMathAlphabet{\mathssf}{OT1}{cmss}{m}{sl}
\DeclareMathAlphabet{\mathcsf}{OT1}{cmss}{sbc}{n}

\newcommand{\randomvalue}[1]{\eurm{\uppercase{#1}}}

\DeclareSymbolFont{bsfletters}{OT1}{cmss}{bx}{n}  
\DeclareSymbolFont{ssfletters}{OT1}{cmss}{m}{n}
\DeclareMathSymbol{\bsfGamma}{0}{bsfletters}{'000}
\DeclareMathSymbol{\ssfGamma}{0}{ssfletters}{'000}
\DeclareMathSymbol{\bsfDelta}{0}{bsfletters}{'001}
\DeclareMathSymbol{\ssfDelta}{0}{ssfletters}{'001}
\DeclareMathSymbol{\bsfTheta}{0}{bsfletters}{'002}
\DeclareMathSymbol{\ssfTheta}{0}{ssfletters}{'002}
\DeclareMathSymbol{\bsfLambda}{0}{bsfletters}{'003}
\DeclareMathSymbol{\ssfLambda}{0}{ssfletters}{'003}
\DeclareMathSymbol{\bsfXi}{0}{bsfletters}{'004}
\DeclareMathSymbol{\ssfXi}{0}{ssfletters}{'004}
\DeclareMathSymbol{\bsfPi}{0}{bsfletters}{'005}
\DeclareMathSymbol{\ssfPi}{0}{ssfletters}{'005}
\DeclareMathSymbol{\bsfSigma}{0}{bsfletters}{'006}
\DeclareMathSymbol{\ssfSigma}{0}{ssfletters}{'006}
\DeclareMathSymbol{\bsfUpsilon}{0}{bsfletters}{'007}
\DeclareMathSymbol{\ssfUpsilon}{0}{ssfletters}{'007}
\DeclareMathSymbol{\bsfPhi}{0}{bsfletters}{'010}
\DeclareMathSymbol{\ssfPhi}{0}{ssfletters}{'010}
\DeclareMathSymbol{\bsfPsi}{0}{bsfletters}{'011}
\DeclareMathSymbol{\ssfPsi}{0}{ssfletters}{'011}
\DeclareMathSymbol{\bsfOmega}{0}{bsfletters}{'012}
\DeclareMathSymbol{\ssfOmega}{0}{ssfletters}{'012}

\newcommand{\rvd}{{\randomvalue{d}}}	

\newcommand{\rvr}{{\randomvalue{r}}}	

\newcommand{\rvt}{{\randomvalue{t}}}	

\newcommand{\rvw}{{\randomvalue{w}}}	

\newcommand{\rvx}{{\randomvalue{x}}}	

\newcommand{\rvy}{{\randomvalue{y}}}	

\newcommand{\rvz}{{\randomvalue{z}}}	

\begin{document}

\title{ORBGRAND Is Almost Capacity-Achieving}
\author{Mengxiao Liu, Yuejun Wei, Zhenyuan Chen, and Wenyi Zhang, \IEEEmembership{Senior~Member, IEEE}
\thanks{M. Liu and W. Zhang are with the CAS Key Laboratory of Wireless-Optical Communications, and the Department of Electronic Engineering and Information Science, University of Science and Technology of China, Hefei, 230027, China; Y. Wei and Z. Chen are with the Wireless Research Department, Huawei Technologies, Shanghai, 201206, China. (corresponding author: Wenyi Zhang, wenyizha@ustc.edu.cn)

The work of M.~Liu and W.~Zhang was supported by the National Key Research and Development Program of China under Grant 2018YFA0701603.
}
}

\maketitle

\begin{abstract}
Decoding via sequentially guessing the error pattern in a received noisy sequence has received attention recently, and ORBGRAND has been proposed as one such decoding algorithm that is capable of utilizing the soft information embedded in the received noisy sequence. An information theoretic study is conducted for ORBGRAND, and it is shown that the achievable rate of ORBGRAND {using independent and identically distributed random codebooks} almost coincides with the channel capacity, for an additive white Gaussian noise channel under antipodal input. For finite-length codes, improved guessing schemes motivated by the information theoretic study are proposed that attain lower error rates than ORBGRAND, especially in the high signal-to-noise ratio regime.
\end{abstract}

\section{Introduction}
\label{sec:intro}

Guessing random additive noise decoding (GRAND) has been recently proposed as a novel decoding approach \cite{duffy19:it}, and has received attention quickly; see, e.g., {\cite{solomon20:icc}--\cite{abbas21:arxiv}}. The basic idea of GRAND can be easily explained for binary symmetric channels. For such channels, given a length-$N$ transmitted codeword $\underline{x}$, the received noisy sequence $\underline{\rvy}$ can be written as the modulo-two sum of $\underline{x}$ and a binary noise sequence $\underline{\rvz}$, i.e., $\underline{\rvy} = \underline{x} \oplus \underline{\rvz}$. Given a sample of $\underline{\rvy}$ as $\underline{y}$, GRAND sequentially enumerates possible noise sequences in descending order of their probabilities, and stops when it identifies the first noise sequence $\underline{z}$ such that $\underline{y} \oplus \underline{z}$ is a valid codeword. That there exists no noise sequence with a larger probability to render a valid codeword is equivalent to that the identified valid codeword by GRAND is the maximum likelihood (ML) decoding result. So GRAND is essentially equivalent to the ML decoder, and is capacity-achieving under capacity-achieving input distribution {(i.e., uniform Bernoulli distribution for binary symmetric channels)}. This argument has been made rigorous in \cite{duffy19:it} for channels with discrete additive noise processes, along with an asymptotic analysis of the error exponent and the guessing complexity. {The feasibility of GRAND has also been testified by chip implementation \cite{riaz22:comsnets}.}

Extending the basic idea of GRAND to more general channels, say, the additive white Gaussian noise (AWGN) channel, is not straightforward. For an AWGN channel, the soft information embedded in the received noisy sequence apparently provides essential help compared with hard-decision channel output, but it is clearly impossible to enumerate all possible real-valued noise sequences. Under this situation, a variant of GRAND, called soft GRAND (SGRAND), has been proposed in \cite{solomon20:icc}. It basically enumerates possible hard-decision error patterns in descending order of their probabilities, and is thus equivalent to the ML decoder.\footnote{{Essentially the same idea has also been briefly mentioned in an earlier work \cite{valembois01:cl}.}} {Since the ordering of hard-decision error patterns depends upon the actual values of channel reliabilities (i.e., absolute values of received noisy sequence), when implementing SGRAND, a sequential online algorithm is required for generating the hard-decision error patterns (see \cite{solomon20:icc} for details), which limits the efficiency of hardware implementation of SGRAND.} {There are also extensions of GRAND and its variants to other channels; see, for example, bursty channels \cite{an22:com} \cite{ercan22:arxiv}, fading channels \cite{sarieddeen22:arxiv} \cite{abbas22:arxiv}, multiple access channels \cite{solomon21:isit}, turbo decoding \cite{galligan22:arxiv}.}

Yet another variant of GRAND, called ordered reliability bits GRAND (ORBGRAND), has been proposed in \cite{duffy21:icassp} \cite{duffy22:tsp}. Similar to SGRAND, ORBGRAND also enumerates possible hard-decision error patterns. {But its utilization of soft information is based upon the ordering relation of channel reliabilities only, without requiring their actual values. Consequently, there exists efficient algorithms based upon partitioning of integers that generates the hard-decision error patterns in a recursive offline fashion; see, e.g., \cite{duffy21:icassp} \cite{duffy22:tsp} \cite{condo21:arxiv}. This fact renders ORBGRAND and its several variants an attractive solution for efficient hardware implementation \cite{condo21:arxiv} \cite{abbas22:vlsi} \cite{condo22:cst}.}

In this paper, we analyze the achievable rate of ORBGRAND for a real AWGN channel under antipodal input. {Noting that the decoding rule of ORBGRAND does not obey the ML criterion (see Section \ref{sec:framework} for details), it is a mismatched decoder. We hence evaluate the generalized mutual information} (GMI) of ORBGRAND, which is a lower bound of the mismatch capacity {when independent and identically distributed (i.i.d.) random codebooks are employed,} and has been extensively used for analyzing mismatched decoding systems \cite{ganti00:it} \cite{lapidoth02:it} \cite{weingarten04:it}. A surprising observation from the GMI analysis is that in the asymptotic regime of large coding block length, there is virtually no gap between the GMI of ORBGRAND and the channel capacity.

That ORBGRAND is almost capacity-achieving can be heuristically explained by considering another closely related variant of GRAND which we call cdf-GRAND, wherein the soft information is based upon the cumulative distribution function (cdf) of the channel reliability values. By inspecting cdf-GRAND, we further propose improved guessing schemes that attain lower error rates than ORBGRAND, for finite-length codes, especially in the high signal-to-noise ratio (SNR) regime.

\section{General Decoding Rule for Guessing Decoders}
\label{sec:framework}

We consider antipodal input over a discrete-time real AWGN channel; that is, the memoryless channel input-output relationship is
\begin{eqnarray}
    \label{eqn:AWGN-channel}
    \rvy = \rvx + \rvz,
\end{eqnarray}
where the channel input $\rvx$ is $\sqrt{P}$ or $-\sqrt{P}$ with equal probability $1/2$, and the channel noise $\rvz \sim \mathcal{N}(0, 1)$. So the SNR is $P$. Denote the probability density function (pdf) of $\rvz$ by $\phi(z)$, i.e.,
\begin{eqnarray}
    \phi(z) = \frac{1}{\sqrt{2\pi}} e^{-\frac{z^2}{2}}, \quad -\infty < z < \infty.
\end{eqnarray}

For coding block length $N$ and coding rate $R$ nats per channel use, there are $M = e^{NR}$ messages, uniformly randomly selected for transmission. For transmitting message $m$, the corresponding codeword is denoted by $\underline{x}(m) = [x_1(m), x_2(m), \ldots, x_N(m)]$. The transmitted codeword induces a received noisy sequence $\underline{\rvy} = [\rvy_1, \rvy_2, \ldots, \rvy_N]$ according to the AWGN channel law (\ref{eqn:AWGN-channel}). For $n = 1, \ldots, N$, denote by $\rvr_n$ the rank of $|\rvy_n|$ among the sorted array consisting of $\{|\rvy_1|, |\rvy_2|, \ldots, |\rvy_N|\}$, from $0$ (the smallest) to $N - 1$ (the largest). We denote by $\Psi(|y|)$ the cdf of $|\rvy|$, given by
\begin{eqnarray}
    \Psi(|y|) = \int_{\sqrt{P} - |y|}^{\sqrt{P} + |y|} \phi(z) \mathrm{d}z.
\end{eqnarray}

For decoding via sequentialy guessing the error pattern in a sample of $\underline{\rvy}$ as $\underline{y}$, we can write the decoding rule in the following general form as
\begin{eqnarray}
    \label{eqn:decoding-rule}
    \widehat{m} &=& \mathrm{arg}\min_{m = 1, \ldots, M} \frac{1}{N} \sum_{n = 1}^N \nonumber\\
    &&\quad\quad \gamma_n(\underline{y}) \cdot \mathbf{1}(\mathrm{sgn}(y_n) \cdot x_n(m) < 0),
\end{eqnarray}
where $\left\{\gamma_n\right\}_{n = 1, \ldots, N}$ are nonnegative functions of $\underline{y}$ to capture the impact of the soft information embedded in the received noisy sequence.\footnote{One may further allow $\left\{\gamma_n\right\}_{n = 1, \ldots, N}$ to depend upon both $\underline{y}$ and $\underline{x}(m)$. Indeed, this is exactly the case for some recent variants of ORBGRAND; see, e.g., \cite{condo21:arxiv} (also see the discussion in Section \ref{sec:improvement}). We do not analyze such generalization in the present paper.} Note that we allow $\gamma_n$ for each $n = 1, \ldots, N$ to depend upon the entire sequence $\underline{y}$. The indicator function $\mathbf{1}(\mathrm{sgn}(y_n) \cdot x_n(m) < 0)$ is equal to one if the sign of $y_n$ is opposite to that of $x_n(m)$, and is zero otherwise; that is, for each potential codeword $\underline{x}(m)$, $m = 1, \ldots, M$, we only keep the positions in $\underline{y}$ where hard-decision errors occur.

{It may appear counterintuitive at first glance that the decoding rule (\ref{eqn:decoding-rule}), which enumerates codewords, is equivalent to guessing decoders, which enumerates hard-decision error patterns. To see why this is the case, let us notice that a guessing decoder executes the following algorithm:}
\begin{itemize}
    \item {Sort the subsets of $\{1, 2, \ldots, N\}$ such that their corresponding partial sums}
    {
    \begin{eqnarray*}
        \mathcal{PS} := \left\{\sum_{i \in \mathcal{I}} \gamma_i(\underline{y}): \mathcal{I} \subseteq \{1, 2, \ldots, N\}\right\}
    \end{eqnarray*}}
    {are in ascending order. Each subset $\mathcal{I}$ corresponds to a hard-decision error pattern.}
    \item {For each subset $\mathcal{I}$ in $\mathcal{PS}$, flip the signs at positions in $\mathrm{sgn}(\underline{y})$ indicated by $\mathcal{I}$. Denote the flipped sequence by $\mathrm{sgn}(\underline{y})_\mathcal{I}$.}
    \item {If the flipped sequence $\mathrm{sgn}(\underline{y})_\mathcal{I}$ is a codeword, stop and declare the codeword as the decoded one; otherwise, draw the next subset $\mathcal{I}$ in $\mathcal{PS}$ and repeat the procedure.}\footnote{{In the worst case, all the subsets in $\mathcal{PS}$ need to be enumerated.} In implementation, this sequential procedure may be terminated after a maximum number of trials if a codeword still cannot be identified, and GRAND with this additional termination rule is called GRAND with abandonment (GRANDAB) \cite{duffy19:it}. {In our theoretical analysis in Section \ref{sec:orbgrand}, we do not consider this termination rule, but in the numerical simulation of finite-length codes in Section \ref{sec:improvement} we will adopt it.}}
\end{itemize}

{If a guessing decoder, executing the above algorithm, stops with a codeword corresponding to message $\hat{m}$, then $\hat{m}$ must also solve (\ref{eqn:decoding-rule}). Otherwise, supposing that it is another message $m'$ rather than $\hat{m}$ that solves (\ref{eqn:decoding-rule}), then in $\mathcal{PS}$ the partial sum associated with $m'$ would be smaller than that associated with $\hat{m}$, and consequently, the guessing decoder would stop with the codeword corresponding to $m'$ rather than $\hat{m}$. This leads to a contradiction.}

We call the sorted subsets of hard-decision error patterns the guessing scheme associated with $\{\gamma_n\}_{n = 1, \ldots, N}$. {Clearly, different choices of $\{\gamma_n\}_{n = 1, \ldots, N}$ correspond to different orderings of subsets of hard-decision error patterns in their guessing decoder implementation.} Some representative examples include:
\begin{itemize}
    \item When $\gamma_n(\underline{y}) = 1$ for all $n = 1, \ldots, N$ and all $\underline{y}$, the decoding rule (\ref{eqn:decoding-rule}) is the original GRAND \cite{duffy19:it}, which simply searches for the valid codeword with the fewest hard-decision errors, without utilizing any soft information.
    \item When $\gamma_n(\underline{y}) = |y_n|$, the decoding rule (\ref{eqn:decoding-rule}) is in fact equivalent to the ML decoding rule, which is capacity-achieving. {Such an equivalence relation} has appeared in \cite[Chap. 10, Sec. 1]{lin04:book}, and for completeness we give a proof in a slightly more general form in Appendix \ref{appendix:SGRAND-optimality}. This decoding rule is also equivalent to SGRAND \cite{solomon20:icc}, which requires a dynamic data structure to generate the guessing scheme in an online fashion.
    \item When $\gamma_n(\underline{y}) = r_n/N$, {where $r_n$ is the realization of the random variable $\rvr_n$, which has been defined as the rank of $|\rvy_n|$ among the sorted array consisting of $\{|\rvy_n|\}_{n = 1, \ldots, N}$ in ascending order,} the decoding rule (\ref{eqn:decoding-rule}) is ORBGRAND \cite{duffy21:icassp}.\footnote{In the literature, the rank is from $1$ to $N$, whereas here we let the rank be from $0$ to $N - 1$ and be further normalized by $N$, for convenience of analysis in Section \ref{sec:orbgrand}. This difference is immaterial as $N$ gets large.}
    \item When $\gamma_n(\underline{y}) = \Psi(|y_n|)$, the cdf of $|\rvy|$ evaluated at $|y_n|$, we call the resulting decoding rule cdf-GRAND. cdf-GRAND turns out to be closely related to ORBGRAND, since both achieve the same GMI, as will be shown in Appendix \ref{appendix:cdf-GRAND}. Similar to SGRAND, cdf-GRAND also requires the decoder to generate the guessing scheme online, so it is not amenable to hardware implementation compared with ORBGRAND. But it provides a convenient way of understanding the behavior of ORBGRAND, and suggests ways of improving ORBGRAND, as will be elaborated in Section \ref{sec:improvement}.
\end{itemize}

{It is worth noting that, since $\{\gamma_n\}_{n = 1, \ldots, N}$ is always equal to $1$ for GRAND and is a permutation of equispaced points $\{1/N, 2/N, \ldots, 1\}$ for ORBGRAND, its discreteness leads to multiple occurrences of ties when picking the decoded message index $m$ in (\ref{eqn:decoding-rule}).} For our analysis of the achievable rate of ORBGRAND in the next section, we focus on the asymptotic regime of $N \rightarrow \infty$ and let ties be broken arbitrarily. For finite-length codes, when several codewords simultaneously attain the minimum in (\ref{eqn:decoding-rule}), a further performance gain can be reaped by adopting a technique proposed in \cite{abbas21:arxiv}; that is, comparing their Euclidean distances to the received noisy sequence $\underline{y}$ and declaring the message whose codeword has the smallest Euclidean distance as the decoded message. In our numerical study in Section \ref{sec:improvement} we adopt this strategy.

\section{Achievable Rate of ORBGRAND}
\label{sec:orbgrand}

Under the general decoding rule (\ref{eqn:decoding-rule}) in the previous section, given $\underline{\rvy}$, ORBGRAND solves the following search problem:
\begin{eqnarray}
    \label{eqn:ORBGRAND}
    \widehat{m} &=& \mathrm{arg}\min_{m = 1, \ldots, M} \rvd(m),\nonumber\\
    \rvd(m) &=& \frac{1}{N} \sum_{n = 1}^N \frac{\rvr_n}{N} \cdot \mathbf{1}(\mathrm{sgn}(\rvy_n) \cdot x_n(m) < 0),
\end{eqnarray}
{where we call $\rvd(m)$ the decoding metric for message $m$.}

We have the following theorem characterizing an achievable rate of ORBGRAND.
\begin{thm}
    \label{thm:ORBGRAND-rate}
    An achievable rate of ORBGRAND (\ref{eqn:ORBGRAND}) under antipodal input over a real AWGN channel (\ref{eqn:AWGN-channel}) is given by
    \begin{eqnarray}
        &&I_\mathrm{ORBGRAND} = \ln 2 - \nonumber\\
        && \inf_{\theta < 0}\Big\{\int_0^1 \ln \left(1 + e^{\theta t}\right) \mathrm{d}t - \theta \int_0^\infty \Psi(t) \phi(t + \sqrt{P}) \mathrm{d}t\Big\}\nonumber\\
    \end{eqnarray}
    in nats/channel use.
\end{thm}
\textit{Proof:} See Section \ref{subsec:orbgrand-proof}.

\subsection{Discussion of $I_\mathrm{ORBGRAND}$ in Theorem \ref{thm:ORBGRAND-rate}}
\label{subsec:orbgrand-discussion}

Our first remark is that $I_\mathrm{ORBGRAND}$ in Theorem \ref{thm:ORBGRAND-rate} is measured in nats rather than bits. In numerical plots it is usually more convenient to convert the unit to bit by scaling.

We numerically compare $I_\mathrm{ORBGRAND}$ with two rates:
\begin{itemize}
    \item The channel capacity of (\ref{eqn:AWGN-channel}) under antipodal input, which is also the rate achieved by SGRAND.
    \item The channel capacity of (\ref{eqn:AWGN-channel}) under hard-decision output $\mathrm{sgn}(\rvy)$ and antipodal input, which is also the rate achieved by GRAND.
\end{itemize}

\begin{figure}[t]
    \centering
    \includegraphics[width=8cm]{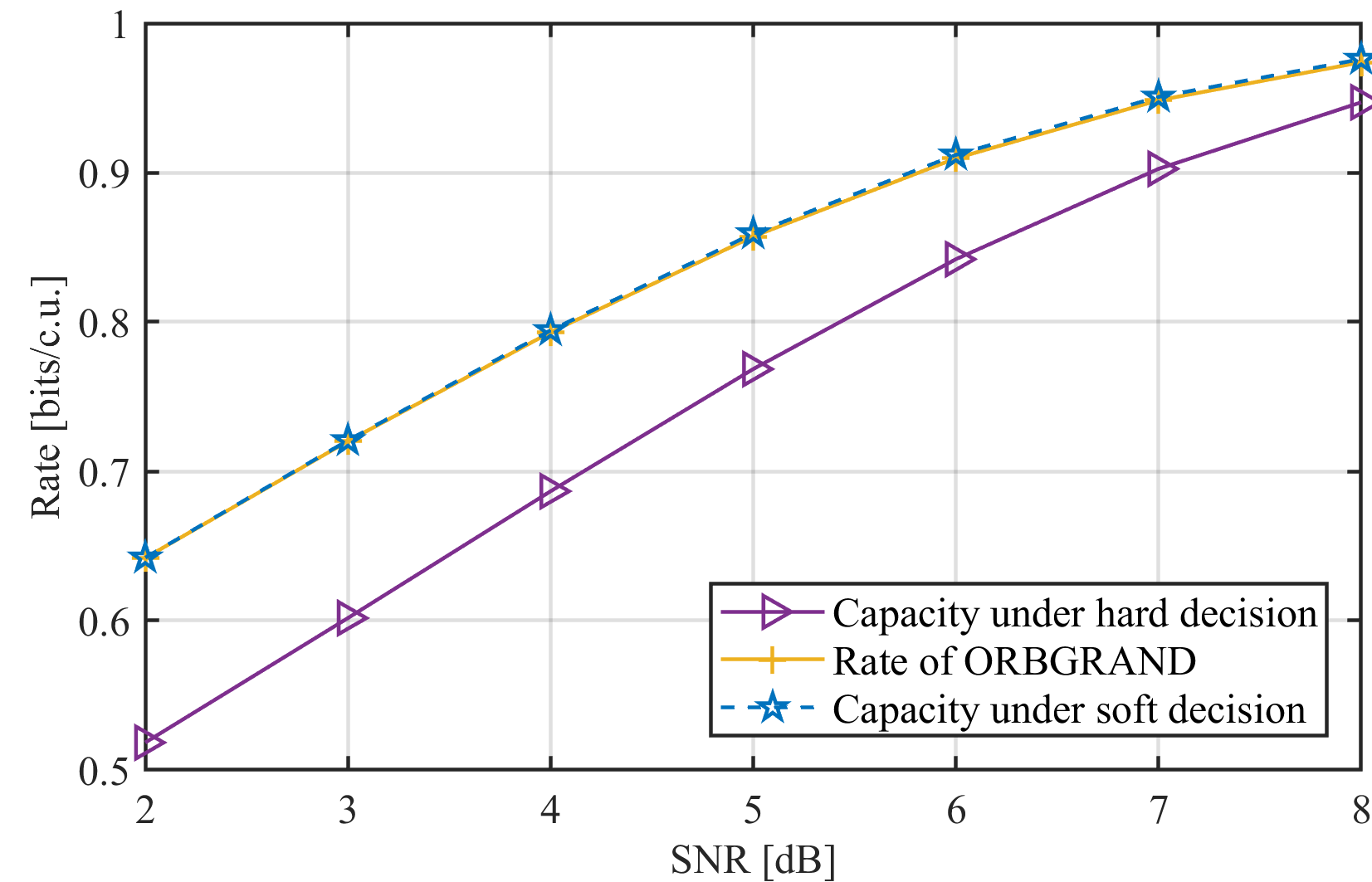}
    \caption{Comparison of $I_\mathrm{ORBGRAND}$ and channel capacity.}
    \label{fig:GMI-capacity} 
\end{figure}

These rates are plotted in Figure \ref{fig:GMI-capacity}. It is surprising to notice that despite the fact that ORBGRAND is not an ML decoding rule, there is virtually no gap between $I_\mathrm{ORBGRAND}$ and the channel capacity. {An explanation of this behavior is provided in Section \ref{sec:improvement}, with the aid of cdf-GRAND.}

Recently, some variants of ORBGRAND have been proposed, and numerical simulations show that they may attain lower error rates for finite-length codes at high SNR \cite{condo21:arxiv} \cite{abbas22:vlsi}. We do not analyze their {achievable} rates. As shown in Figure \ref{fig:GMI-capacity}, $I_\mathrm{ORBGRAND}$ is almost capacity-achieving already, so the room for improvement in terms of achievable rate will be rather limited.

\subsection{Proof of Theorem \ref{thm:ORBGRAND-rate}}
\label{subsec:orbgrand-proof}

Our proof is essentially a calculation of the GMI of ORBGRAND (\ref{eqn:ORBGRAND}).

We consider the ensemble of codebooks consisting of i.i.d. codewords; that is, each codeword
\begin{eqnarray*}
\underline{\rvx}(m) = [\rvx_1(m), \rvx_2(m), \ldots, \rvx_N(m)]
\end{eqnarray*}
consists of i.i.d. samples of $\rvx$, and the $M$ codewords are pairwise independent.

We analyze the average decoding error probability, further averaged over the i.i.d. codebook ensemble. Due to the i.i.d. nature of the codebook ensemble, the average decoding error probability is equal to the decoding error probability conditioned upon that $m = 1$ is the sent message.

A fact that will be useful for our subsequent derivation is the following lemma regarding the resulting $\{|\rvy_n|\}_{n = 1, \ldots, N}$.
\begin{lem}
    \label{lem:independent-absY}
    Under the i.i.d. codebook ensemble, the resulting $\{|\rvy_n|\}_{n = 1, \ldots, N}$ are i.i.d. with cdf $\Psi(|y|)$.
\end{lem}
\textit{Proof:} This is an exercise of elementary probability theory, noting that the antipodal input alphabet is symmetric with respect to the origin and so is the pdf of the AWGN $\rvz$. \textbf{Q.E.D.}

We have the following three lemmas that characterize the asymptotic behavior of $\rvd(1)$ and $\rvd(m')$, $m' \neq 1$, which are random variables induced by the i.i.d. codebook and the AWGN channel, appearing in (\ref{eqn:ORBGRAND}).

\begin{lem}
    \label{lem:ED1}
    Assuming that $m = 1$ is the sent message, we have
    \begin{eqnarray}
        \label{eqn:lem-ED1}
        \lim_{N \rightarrow \infty} \mathbb{E} \rvd(1) = \int_0^\infty \Psi(t) \phi(t + \sqrt{P}) \mathrm{d}t.
    \end{eqnarray}
\end{lem}
\textit{Proof:} See Appendix \ref{appendix:ED1}.

\begin{lem}
    \label{lem:varD1}
    Assuming that $m = 1$ is the sent message, we have
    \begin{eqnarray}
        \label{eqn:lem-varD1}
        \lim_{N \rightarrow \infty} \mathrm{var} \rvd(1) = 0.
    \end{eqnarray}
\end{lem}
\textit{Proof:} See Appendix \ref{appendix:varD1}.

\begin{lem}
    \label{lem:EDm}
    Assuming that $m = 1$ is the sent message, for any $m' \neq 1$ and any $\theta < 0$, {we have the following behavior of the asymptotic logarithmic conditional moment generating function of $\rvd(m')$,} almost surely,
    \begin{eqnarray}
        \Lambda(\theta) &:=& \lim_{N \rightarrow \infty} \frac{1}{N} \ln \mathbb{E}\left\{e^{N \theta \rvd(m')} \Big| \underline{\rvy}\right\}\nonumber\\
        &=& \int_0^1 \ln (1 + e^{\theta t}) \mathrm{d}t - \ln 2.
    \end{eqnarray}
\end{lem}
\textit{Proof:} See Appendix \ref{appendix:EDm}.

For any $\delta > 0$, define event
\begin{eqnarray*}
\mathcal{A}_\delta = \left\{\rvd(1) \geq \int_0^\infty \Psi(t) \phi(t + \sqrt{P}) \mathrm{d}t + \delta\right\};
\end{eqnarray*}
{that is, the decoding metric for the sent message, i.e., $\rvd(1)$, exceeds its mean (see Lemma \ref{lem:ED1}) by an amount of $\delta$.} The average decoding error probability can be written as
\begin{eqnarray}
    &&\mathrm{Pr}\left[\widehat{m} \neq 1\right]\nonumber\\
    &=& \mathrm{Pr}\left[\widehat{m} \neq 1\big| \mathcal{A}_\delta\right] \mathrm{Pr}\left[\mathcal{A}_\delta\right] + \mathrm{Pr}\left[\widehat{m} \neq 1\big| \mathcal{A}_\delta^c\right] \mathrm{Pr}\left[\mathcal{A}_\delta^c\right]\nonumber\\
    &\leq& \mathrm{Pr}\left[\mathcal{A}_\delta\right] + \mathrm{Pr}\left[\widehat{m} \neq 1\big| \mathcal{A}_\delta^c\right].
\end{eqnarray}

Based upon Lemmas \ref{lem:ED1} and \ref{lem:varD1}, an application of Chebyshev's inequality indicates that for any $\delta > 0$,
\begin{eqnarray}
    \lim_{N \rightarrow \infty} \mathrm{Pr}\left[\rvd(1) \geq \int_0^\infty \Psi(t) \phi(t + \sqrt{P}) \mathrm{d}t + \delta\right] = 0;
\end{eqnarray}
that is, $\mathrm{Pr}\left[\mathcal{A}_\delta\right]$ can be made arbitrarily close to zero for sufficiently large $N$.

On the other hand, due to the decoding rule and the union bound,
\begin{eqnarray}
    \label{eqn:ave-error-prob}
    &&\mathrm{Pr}\left[\widehat{m} \neq 1\big| \mathcal{A}_\delta^c\right] \nonumber\\
    &\leq&
    \mathrm{Pr}\left[\exists m' \neq 1, \rvd(m') < \int_0^\infty \Psi(t) \phi(t + \sqrt{P}) \mathrm{d}t + \delta\right]\nonumber\\
    &\leq& e^{NR} \mathrm{Pr}\left[\rvd(m') < \int_0^\infty \Psi(t) \phi(t + \sqrt{P}) \mathrm{d}t + \delta\right].
\end{eqnarray}
So it suffices to investigate the exponential rate at which $\mathrm{Pr}\left[\rvd(m') < \int_0^\infty \Psi(t) \phi(t + \sqrt{P}) \mathrm{d}t + \delta\right]$ tends towards zero. For this, we apply the law of total expectation to consider (\ref{eqn:Dmprime}).
\begin{figure*}
\begin{eqnarray}\label{eqn:Dmprime}
    \mathrm{Pr}\left[\rvd(m') < \int_0^\infty \Psi(t) \phi(t + \sqrt{P}) \mathrm{d}t + \delta\right] = \mathbb{E} \left[\mathrm{Pr}\left[\rvd(m') < \int_0^\infty \Psi(t) \phi(t + \sqrt{P}) \mathrm{d}t + \delta\Big| \underline{\rvy}\right] \right].
\end{eqnarray}
\end{figure*}

Then an application of Chernoff's bound indicates that for any $N$ and any $\theta < 0$, the inequality (\ref{eqn:chernoff}) holds.
\begin{figure*}
\begin{eqnarray}\label{eqn:chernoff}
    -\frac{1}{N} \ln \mathrm{Pr}\left[\rvd(m') < \int_0^\infty \Psi(t) \phi(t + \sqrt{P}) \mathrm{d}t + \delta \Big| \underline{\rvy}\right]
    \geq \theta \left[\int_0^\infty \Psi(t) \phi(t + \sqrt{P}) \mathrm{d}t + \delta\right] - \frac{1}{N} \ln \mathbb{E}\left\{e^{N \theta \rvd(m')} \Big| \underline{\rvy}\right\}.
\end{eqnarray}
\end{figure*}

Letting $\delta \rightarrow 0$ and $N \rightarrow \infty$, and applying the almost sure limit in Lemma \ref{lem:EDm}, we thus have (\ref{eqn:chernoff-limit}).
\begin{figure*}
\begin{eqnarray}\label{eqn:chernoff-limit}
    \lim_{\delta \rightarrow 0} \lim_{N \rightarrow \infty} -\frac{1}{N} \ln \mathrm{Pr}\left[\rvd(m') < \int_0^\infty \Psi(t) \phi(t + \sqrt{P}) \mathrm{d}t + \delta\right] 
    \geq \theta \int_0^\infty \Psi(t) \phi(t + \sqrt{P}) \mathrm{d}t - \int_0^1 \ln (1 + e^{\theta t}) \mathrm{d}t + \ln 2.
\end{eqnarray}
\end{figure*}

{So for any given $\theta$, for sufficiently small $\delta$ and sufficiently large $N$, the bound (\ref{eqn:Dmprime-bound}) holds.}
\begin{figure*}
\begin{eqnarray}\label{eqn:Dmprime-bound}
    \mathrm{Pr}\left[\rvd(m') < \int_0^\infty \Psi(t) \phi(t + \sqrt{P}) \mathrm{d}t + \delta\right]
    \leq \exp \left\{- N \left[\theta \int_0^\infty \Psi(t) \phi(t + \sqrt{P}) \mathrm{d}t - \int_0^1 \ln (1 + e^{\theta t}) \mathrm{d}t + \ln 2\right]\right\}.
\end{eqnarray}
\end{figure*}

{Plugging this into the right hand side of (\ref{eqn:ave-error-prob}), and optimizing over $\theta$, we have that for any $R$ satisfying (\ref{eqn:R-bound}),}
\begin{figure*}
\begin{eqnarray}\label{eqn:R-bound}
    R &<& \sup_{\theta < 0} \left\{\theta \int_0^\infty \Psi(t) \phi(t + \sqrt{P}) \mathrm{d}t - \int_0^1 \ln (1 + e^{\theta t}) \mathrm{d}t + \ln 2\right\}\nonumber\\
    &=& \ln 2 - \inf_{\theta < 0}\left\{\int_0^1 \ln (1 + e^{\theta t}) \mathrm{d}t - \theta \int_0^\infty \Psi(t) \phi(t + \sqrt{P}) \mathrm{d}t\right\},
\end{eqnarray}
\end{figure*}
the average decoding error probability is guaranteed to vanish asymptotically as $N \rightarrow \infty$. This establishes Theorem \ref{thm:ORBGRAND-rate}.

\section{Improved Guessing Schemes}
\label{sec:improvement}

In order to further understand the behavior of ORBGRAND, it is useful to examine cdf-GRAND, which has $\gamma_n(\underline{y}) = \Psi(|y_n|)$ for $n = 1, \ldots, N$. For sufficiently large $N$, the rank of $|\rvy_n|$ among the sorted array consisting of $\{|\rvy_1|, |\rvy_2|, \ldots, |\rvy_N|\}$, after scaling by $N$, is close to $\Psi(|\rvy_n|)$ with high probability. So intuitively, as $N \rightarrow \infty$, the behavior of ORBGRAND tends to be ``similar'' to that of cdf-GRAND. This intuition is formalized by the following result.

\begin{prop}
    \label{prop:cdf-GRAND-equivalence}
    The GMI of cdf-GRAND is the same as $I_\mathrm{ORBGRAND}$.
\end{prop}
\textit{Proof:} See Appendix \ref{appendix:cdf-GRAND}. \textbf{Q.E.D.}

Compared with ORBGRAND, cdf-GRAND has $\gamma_n(\underline{y}) = \Psi(|y_n|)$ which depends upon $y_n$ only, rather than the entire received noisy sequence $\underline{y}$. This property makes cdf-GRAND not amenable to implementation, but makes it easier to understand compared with ORBGRAND. So by examining cdf-GRAND, we may deepen our understanding of ORBGRAND, and may further design improved guessing schemes.

\begin{figure}[ht]
    \centering
    \includegraphics[width=8cm]{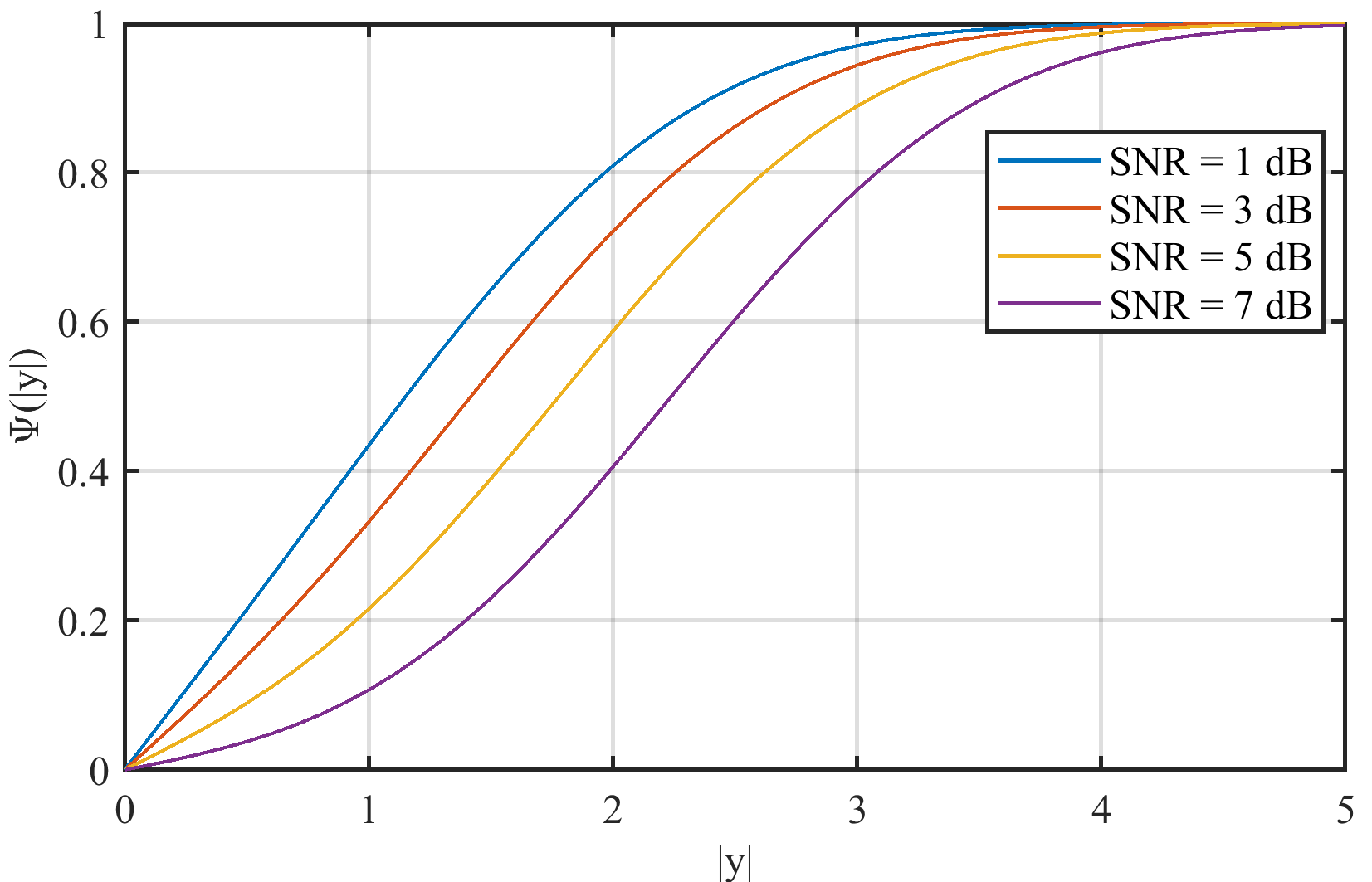}
    \caption{Plots of cdf curves for $|\rvy|$ under different SNRs.}
    \label{fig:CDF-SNR}
\end{figure}

We plot the cdf curves $\Psi(|y|)$ versus $|y|$ in Figure \ref{fig:CDF-SNR}, under different values of the SNR. At low or moderate SNR, the cdf curve $\Psi(|y|)$ has a nearly linear trend for small and moderate values of $|y|$, and saturates for large values of $|y|$. This helps explain why $I_\mathrm{ORBGRAND}$ almost coincides with the channel capacity in the low and moderate SNR regimes in Figure \ref{fig:GMI-capacity}, because a nearly linear $\Psi(|y|)$ is almost equivalent to $|y|$.

At high SNR, the cdf curve $\Psi(|y|)$ has a visible deviation from linearity for small values of $|y|$. This is because when $P \gg 1$, $|\rvy|$ tends to be concentrated around $P$, away from the origin. Although this deviation still does not significantly affect the achievable rate $I_\mathrm{ORBGRAND}$, it leaves room for improvement when considering the error rate performance of finite-length codes.

We propose two improved guessing schemes. The basic idea is to fit the optimal $\gamma_n(\underline{y}) = |y_n|$ more accurately than ORBGRAND and cdf-GRAND, particularly at high SNR. To facilitate implementation, we restrict the guessing schemes to depend upon the rank statistics of $\{|\rvy_n|\}_{n = 1, \ldots, N}$, $\{\rvr_n\}_{n = 1, \ldots, N}$, similar to ORBGRAND, so that they can be generated in an offline fashion. The proposed guessing schemes are mainly for illustrating how ORBGRAND may be improved by examining Figure \ref{fig:CDF-SNR}; we have not exhaustively searched the design space and hence there may well be guessing schemes that perform even better.

\begin{itemize}
    \item Uniform partitioning based ORBGRAND (UP-ORBGRAND): Suppose that we partition the x-axis of Figure \ref{fig:CDF-SNR} into consecutive equal intervals of length $\tau$ (terminated at a sufficiently large $|y|$), and let $\gamma_n(\underline{y}) = i$ if $|y_n|$ falls into the $i$-th interval $[i \tau, (i + 1) \tau)$, for $i = 0, 1, \ldots$. This provides a staircase-like approximation of the optimal $\gamma_n(\underline{y}) = |y_n|$. In order to obtain a guessing scheme that depends upon the rank statistics $\{\rvr_n\}_{n = 1, \ldots, N}$ only, we let $\gamma_n(\underline{y}) = i$ if $\Psi^{-1}(r_n/N)$ falls into the $i$-th interval, where $\Psi^{-1}$ denotes the inverse function of $\Psi$. Due to the deviation of the empirical distribution of the ranks from their true distribution, a finer partitioning may not always lead to better performance than a coarser one, and in general the optimal interval length $\tau$ depends upon the specific code and the SNR.
    \item Biased ORBGRAND (B-ORBGRAND): The idea here is that when the SNR is high, we may simply ignore the CDF curve for small values of $|y|$ in Figure \ref{fig:CDF-SNR}. The portion of the CDF curve for moderate values of $|y|$ still has a good linearity, so ORBGRAND would work well there. In order to realize this idea, it suffices to introduce a bias to ORBGRAND, and consequently let $\gamma_n(\underline{y}) = (r_n + \beta)/N$. The optimal value of the bias $\beta$ depends upon the specific code and the SNR in general.
\end{itemize}

As remarked in footnote 1, there can be more general guessing schemes beyond (\ref{eqn:decoding-rule}), by allowing $\{\gamma_n\}_{n = 1, \ldots, N}$ to depend upon both $\underline{y}$ and $\underline{x}(m)$. For example, the guessing scheme proposed recently in \cite[Eqn. (2)]{condo21:arxiv} can be written as
\begin{eqnarray}
    \label{eqn:iLWO}
    \widehat{m} &=& \mathrm{arg}\min_{m = 1, \ldots, M} \frac{1}{N} \sum_{n = 1}^N \nonumber\\
    &&\quad \frac{r_n v_n}{N} \cdot \mathbf{1}(\mathrm{sgn}(y_n)\cdot x_n(m) < 0),
\end{eqnarray}
where $r_n$ is the rank of $|y_n|$ among the sorted array consisting of $\{|y_n|\}_{n = 1, \ldots, N}$ in ascending order, the same as that in ORBGRAND, and the additional factor $v_n$ is the rank of $|y_n|$ among the subarray of the above sorted $\{|y_n|\}_{n = 1, \ldots, N}$ further satisfying the condition $\mathrm{sgn}(y_n)\cdot x_n(m) < 0$. As will be shown in the following numerical simulation results, the guessing scheme (\ref{eqn:iLWO}) also achieves evident performance improvement compared with ORBGRAND. Its analysis is beyond the scope of this paper, and is an interesting topic left for future research.

In our numerical simulation, we consider two codes: BCH(127, 113) and polar(128, 114),\footnote{There are $6$ CRC bits appended to the $114$ information bits, and therefore in total $120$ bits are fed into the polar encoder. After guessing decoding of the polar code, those codewords not passing the CRC are then detected and discarded.} both with rates close to 0.9 bits per channel use. The design parameters $\tau$ and $\beta$ for these codes at different values of SNR are listed in Table I. As already remarked in footnote 2, the guessing scheme is terminated after a maximum number of trials, denoted by $Q$.

\begin{table*}[ht]
\label{table:parameters}
\begin{center}
    \caption{Design parameters used in numerical simulations}
    \begin{tabular}{|l|l|l|}
    \hline
    SNR & $\tau$ & $\beta$\\
    \hline
    4dB & $0.0452$ & $4$ \\
    \hline
    5dB & $0.079$ & $5$ \\
    \hline
    6dB & $0.1666$ & $6$ \\
    \hline
    7dB & $0.365$ & $8$ \\
    \hline
    \end{tabular}
\end{center}
\end{table*}

\begin{figure}[ht]
    \centering
    \includegraphics[width=8cm]{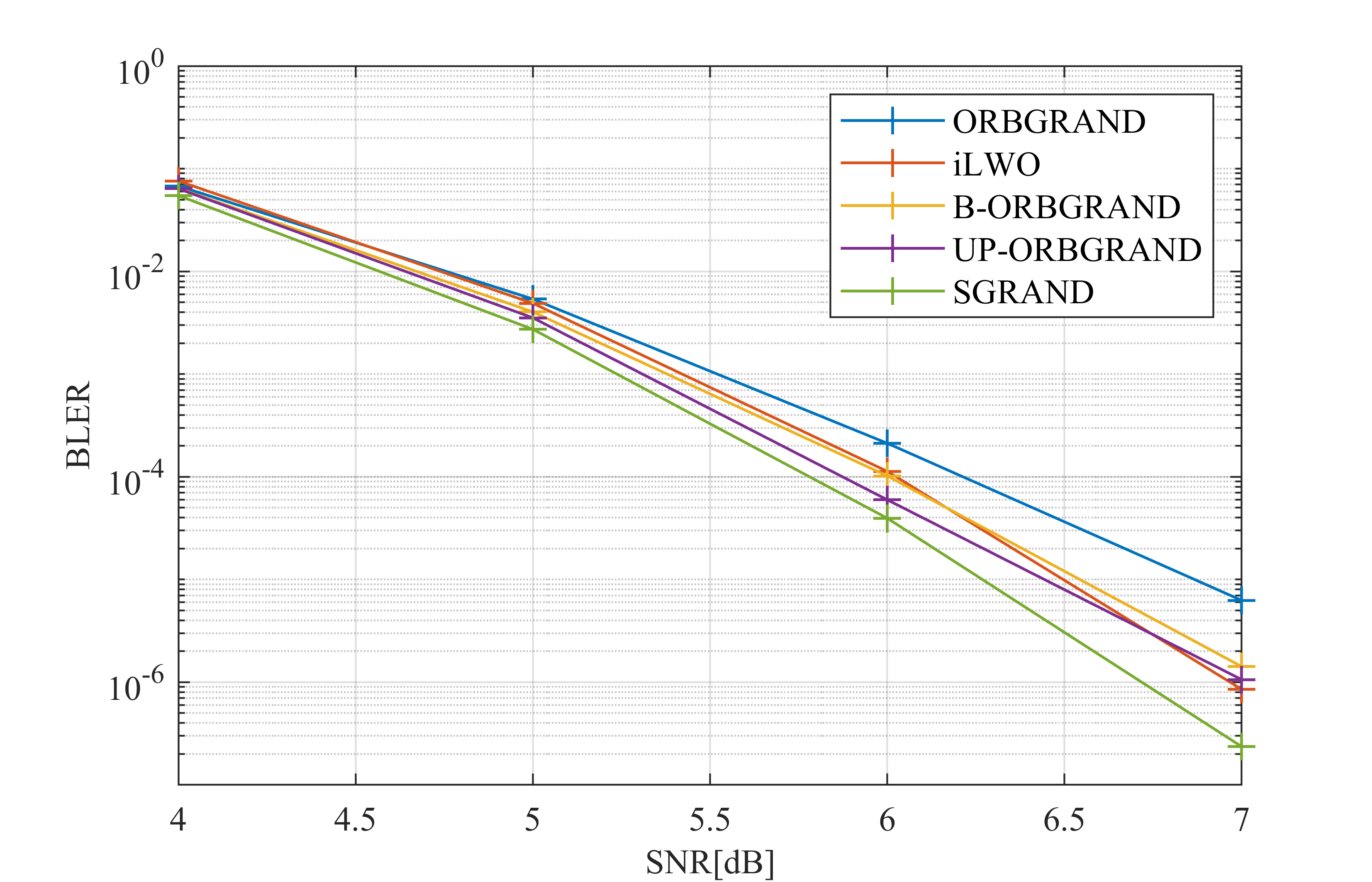}
    \caption{Block error rate performance for BCH(127, 113), $Q = 10^4$.}
    \label{fig:BLER-BCH}
\end{figure}

\begin{figure}[ht]
    \centering
    \includegraphics[width=8cm]{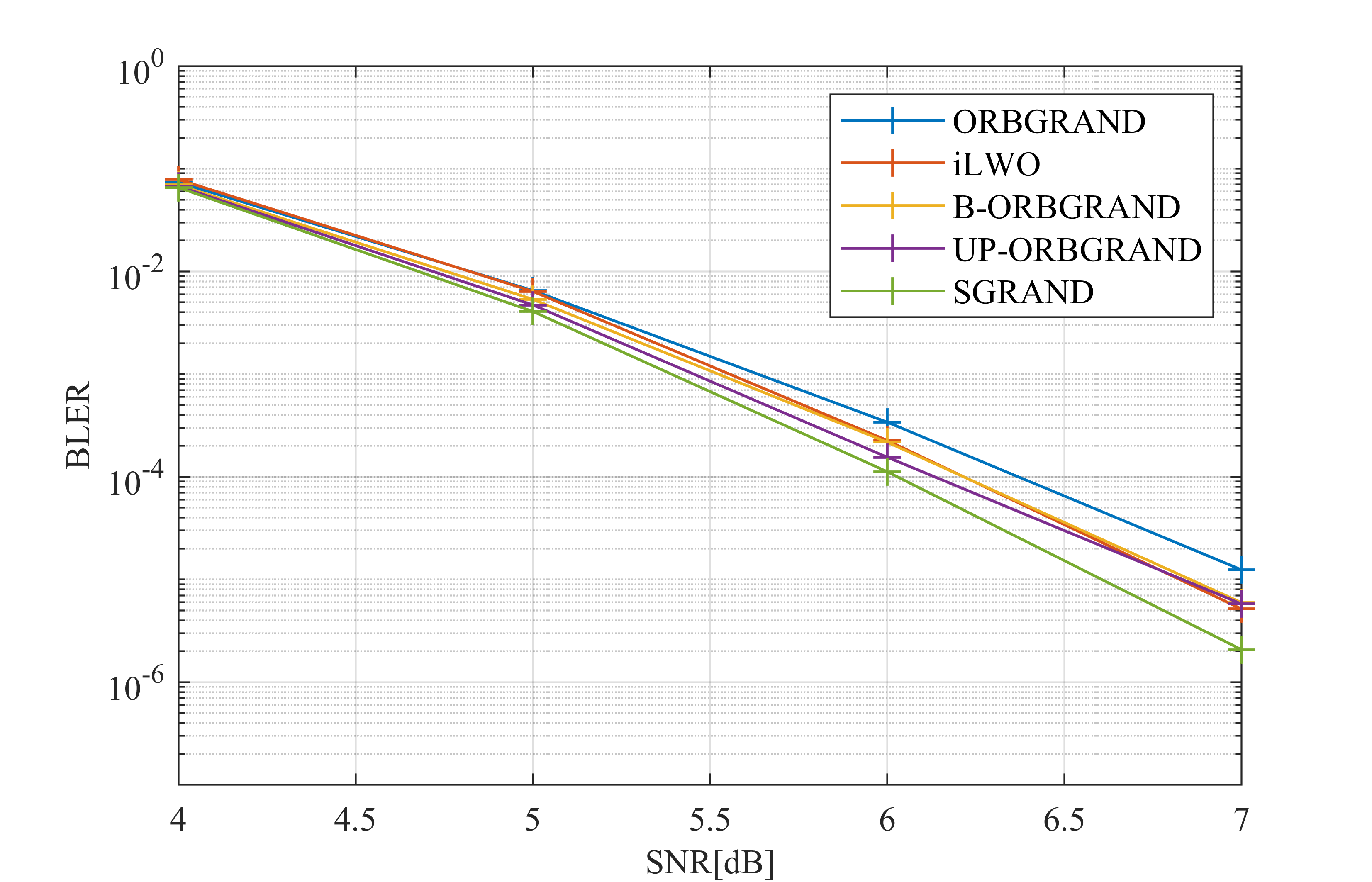}
    \caption{Block error rate performance for Polar(128, 114), $Q = 10^4$.}
    \label{fig:BLER-Polar}
\end{figure}

Figures \ref{fig:BLER-BCH} and \ref{fig:BLER-Polar} display the block error rate (BLER) performance for BCH(127, 113) and Polar(128, 114), respectively, where we fix $Q = 10^4$. We can observe that both UP-ORBGRAND and B-ORBGRAND attain lower BLER compared with ORBGRAND. There still exists quite some gap to SGRAND, and the guessing scheme (\ref{eqn:iLWO}) proposed in \cite{condo21:arxiv} also performs well and may even outperform UP-ORBGRAND and B-ORBGRAND as SNR increases. These facts imply that there is still room for further improvements, left for future research.

We also list mean and variance of the number of guesses for BCH(127, 113) and polar(128, 114) in Tables II and III, respectively. Both mean and variance decrease rapidly with the SNR. It should be noted that although SGRAND has the smallest mean and variance among the considered guessing schemes, the complexity of performing guesses for SGRAND is much higher than that for the other four. This is because as remarked in Section \ref{sec:framework}, SGRAND requires a dynamic data structure to generate the sorted subsets of hard-decision error patterns, in an online fashion.

\begin{table*}[ht]
    \label{table:stat-BCH}
    \begin{center}
        \caption{Mean/variance of the number of guesses for BCH(127, 113), $Q = 10^4$.}
        \begin{tabular}{|l|l|l|l|l|}
        \hline
        SNR & 4dB & 5dB & 6dB & 7dB\\
        \hline
        ORBGRAND & $799.7$ / $4.54\times 10^6$ & $88.5$ / $3.77\times 10^5$ & $7.30$ / $1.52\times 10^4$ & $1.52$ / $5.1\times 10^2$ \\
        \hline
        B-ORBGRAND & $751.3$ / $4.11\times 10^6$ & $74.1$ / $2.69\times 10^5$ & $5.83$ / $5.74\times 10^3$ & $1.49$ / $66.7$ \\
        \hline
        UP-ORBGRAND & $749.2$ / $4.12\times 10^6$ & $69.9$ / $2.43\times 10^5$ & $5.32$ / $3.89\times 10^3$ & $1.52$ / $85.0$ \\
        \hline
        SGRAND & $641.6$ / $3.40\times 10^6$ & $53.4$ / $1.59\times 10^5$ & $3.90$ / $1.38\times 10^3$ & $1.33$ / $5.42$ \\
        \hline
        iLWO (\ref{eqn:iLWO}) \cite{condo21:arxiv} & $865.1$ / $4.90 \times 10^6$ & $81.1$ / $3.12 \times 10^5$ & $5.68$ / $6.08\times 10^3$ & $1.41$ / $36.8$ \\
        \hline
        \end{tabular}
    \end{center}
\end{table*}

\begin{table*}[ht]
    \label{table:stat-polar}
    \begin{center}
        \caption{Mean/variance of the number of guesses for polar(128, 114), $Q = 10^4$.}
        \begin{tabular}{|l|l|l|l|l|}
        \hline
        SNR & 4dB & 5dB & 6dB & 7dB\\
        \hline
        ORBGRAND & $824.8$ / $4.62\times 10^6$ & $90.7$ / $3.99\times 10^5$ & $7.23$ / $1.43\times 10^4$ & $1.51$ / $4.26\times 10^2$ \\
        \hline
        B-ORBGRAND & $789.2$ / $4.38\times 10^6$ & $76.4$ / $2.83\times 10^5$ & $5.79$ / $5.07\times 10^3$ & $1.49$ / $54.6$ \\
        \hline
        UP-ORBGRAND & $764.7$ / $4.15\times 10^6$ & $70.9$ / $2.39\times 10^5$ & $5.37$ / $3.85\times 10^3$ & $1.52$ / $87.2$ \\
        \hline
        SGRAND & $670.8$ / $3.62\times 10^6$ & $53.4$ / $1.68\times 10^5$ & $3.95$ / $1.39\times 10^3$ & $1.33$ / $4.37$ \\
        \hline
        iLWO (\ref{eqn:iLWO}) \cite{condo21:arxiv} & $865.2$ / $4.88 \times 10^6$ & $81.4$ / $3.12 \times 10^5$ & $5.64$ / $5.29\times 10^3$ & $1.41$ / $28.1$ \\
        \hline
        \end{tabular}
    \end{center}
\end{table*}

\begin{figure}[ht]
    \centering
    \includegraphics[width=8cm]{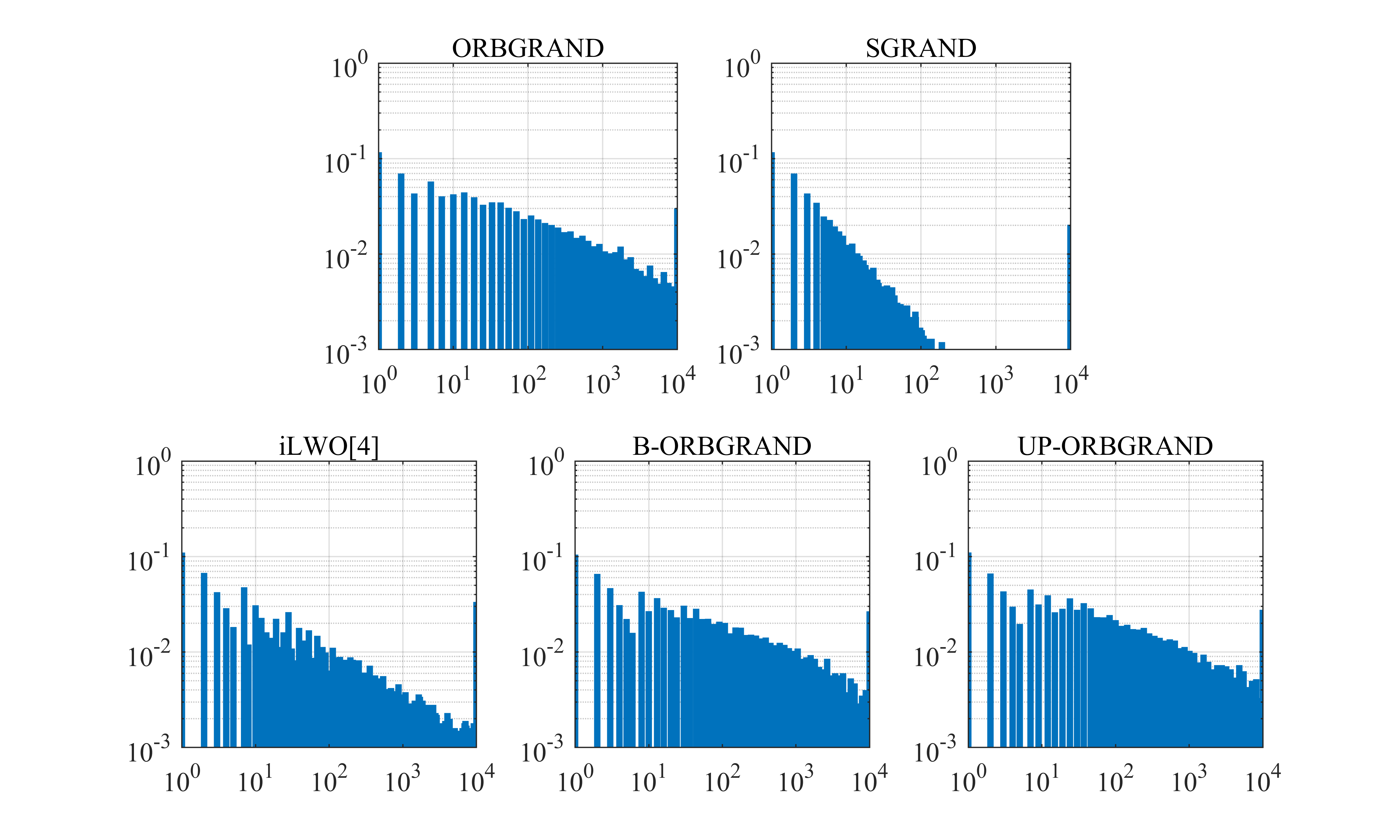}
    \caption{Histogram of guesses for BCH(127, 113), SNR $= 4$dB and $Q = 10^4$.}
    \label{fig:hist-4dB}
\end{figure}

\begin{figure}[ht]
    \centering
    \includegraphics[width=8cm]{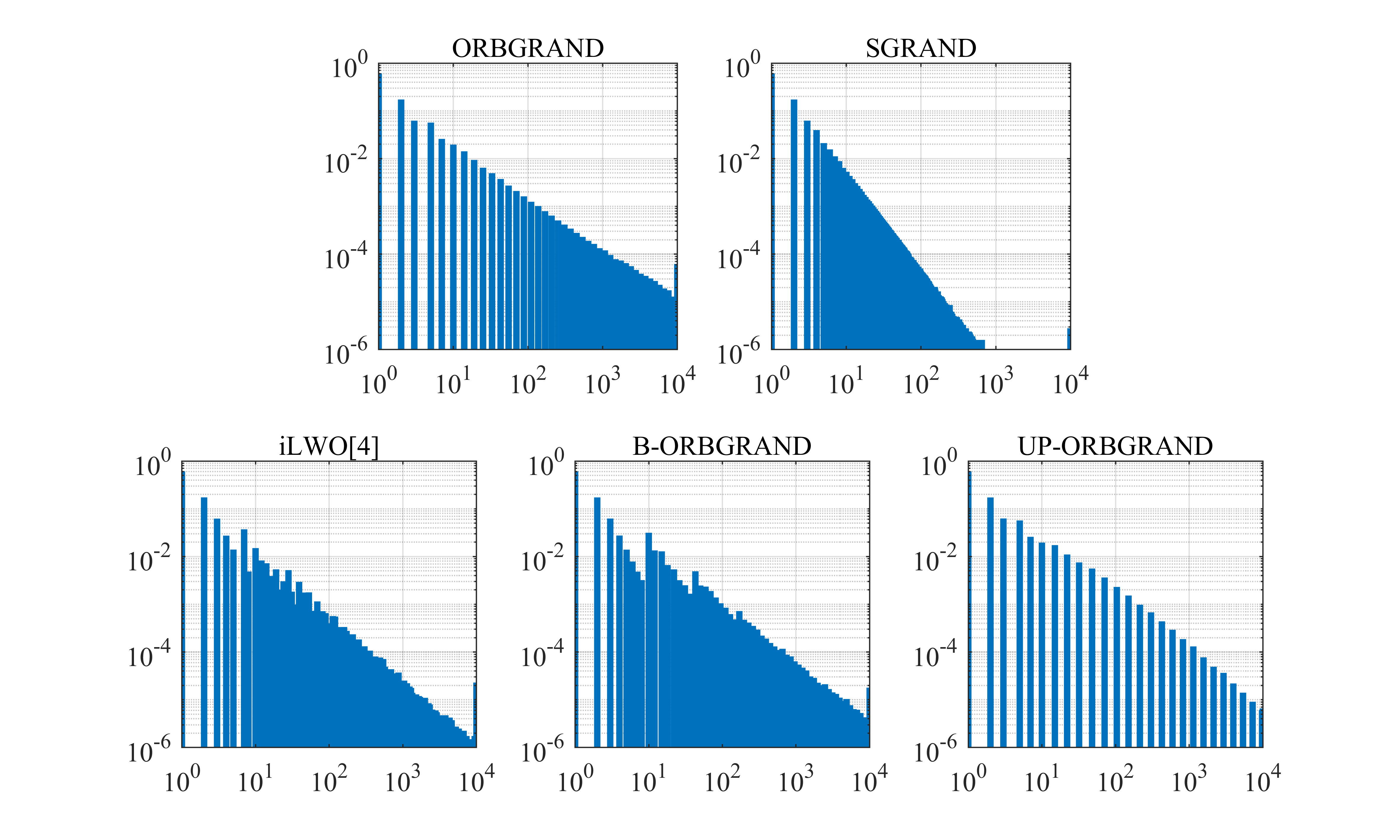}
    \caption{Histogram of guesses for BCH(127, 113), SNR $= 6$dB and $Q = 10^4$.}
    \label{fig:hist-6dB}
\end{figure}

That the stardard deviation (i.e., square root of variance) is almost always much larger than the mean in Tables II and III suggests that the number of guesses may have a long tail in distribution. The empirical histograms displayed in Figures \ref{fig:hist-4dB} and \ref{fig:hist-6dB} confirm this anticipation.

\begin{figure}[ht]
    \centering
    \includegraphics[width=8cm]{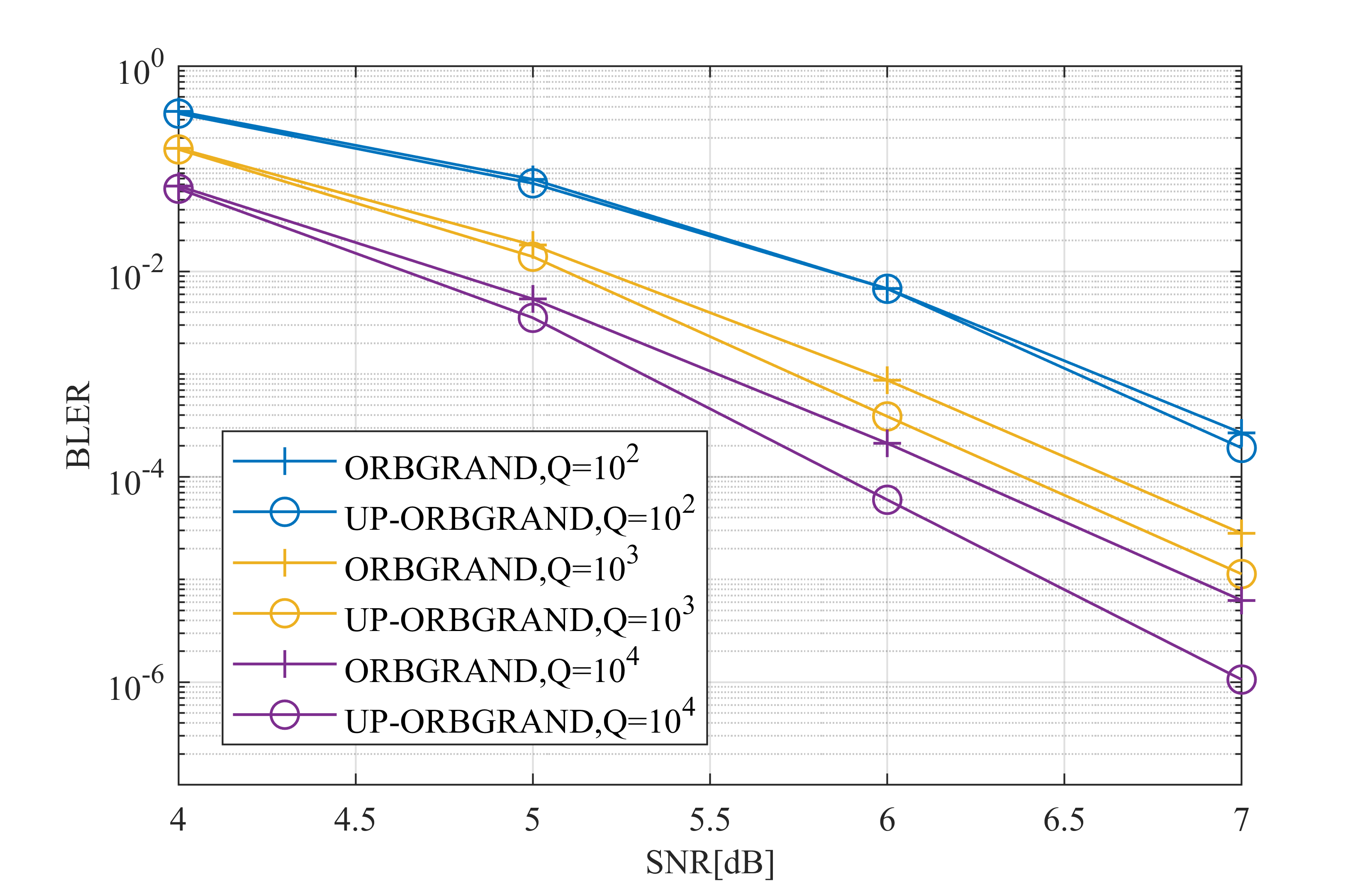}
    \caption{Block error rate performance for BCH(127, 113), under different values of $Q$.}
    \label{fig:Q}
\end{figure}

Finally, Figure \ref{fig:Q} examines the impact of the choice of $Q$, where for clarity we only compare ORBGRAND and UP-ORBGRAND. We observe that although the means listed in Table II are much smaller than $Q$, increasing $Q$ from $10^2$ to $10^4$ still {makes} evident performance improvement. This is exactly due to the long tail propery of the distribution of the number of guesses.

\section{Conclusion}
\label{sec:conclusion}

Based upon a unified general form of guessing decoders and leveraging information-theoretic tools from mismatched decoding, we have calculated an achievable rate of ORBGRAND for antipodal input over the AWGN channel. It is surprising that ORBGRAND achieves a rate almost identical to the channel capacity, despite the fact that it is mismatched to the ML decoder. Motivated by the information-theoretic study, we have also proposed improved guessing schemes that are capable of attaining lower error rates than ORBGRAND, especially in the high SNR regime. The theoretical findings and associated design improvements may be useful for deepening our understanding of guessing decoders, and may be relevant for applications such as ultra-reliable low latency communications. An interesting topic for future research is to explore the ultimate performance limit of guessing decoders and compare them with other universal decoding methods such as the ordered statistics decoding (OSD) algorithm \cite{fossorier95:it} \cite{papadopoulou21:sips}.

\section{Appendix}

\subsection{{Proof of the Equivalence between $r_n(\underline{y}) = |y_n|$ in (\ref{eqn:decoding-rule}) and ML Decoding}}
\label{appendix:SGRAND-optimality}

Here we prove a slightly more general result, and the optimality of $r_n(\underline{y}) = |y_n|$ in (\ref{eqn:decoding-rule}) follows as a corollary.

\begin{lem}
    \label{lem:optimality-decoding-rule}
    Consider a codebook $\{\underline{x}(m)\}_{m = 1, \ldots, M}$ of coding block length $N$ and coding rate $R$, used for a memoryless channel consisting of two possible input letters $1$ and $-1$, and whose output $\rvy$ has conditional probability distributions $q_0(y)$ if $x = 1$ and $q_1(y)$ if $x = -1$, respectively. The following two decoding rules:
    \begin{eqnarray}
        \label{eqn:decoding-rule-form-1}
        \widehat{m} &=& \mathrm{arg}\min_{m = 1, \ldots, M} \frac{1}{N}\sum_{n = 1}^N \left|\ln \frac{q_0(y_n)}{q_1(y_n)}\right| \nonumber\\
        && \cdot \mathbf{1}\left(\mathrm{sgn}\left(\ln \frac{q_0(y_n)}{q_1(y_n)}\right) x_n(m) < 0\right),\\
        \label{eqn:decoding-rule-form-2}
        \mbox{and}\; \widehat{m} &=& \mathrm{arg}\max_{m = 1, \ldots, M} \frac{1}{N}\sum_{n = 1}^N \ln p(y_n |x_n(m)),
    \end{eqnarray}
    are equivalent.
\end{lem}
\textit{Proof:} The sum in (\ref{eqn:decoding-rule-form-1}) can be expanded as (\ref{eqn:decoding-rule-form-1-proof}),
\begin{figure*}
\begin{eqnarray}
    \label{eqn:decoding-rule-form-1-proof}
    &&\sum_{n = 1}^N \left|\ln \frac{q_0(y_n)}{q_1(y_n)}\right| \cdot \mathbf{1}\left(\mathrm{sgn}\left(\ln \frac{q_0(y_n)}{q_1(y_n)}\right) x_n(m) < 0\right)\nonumber\\
    &=& \sum_{n: x_n(m) = 1, q_0(y_n) < q_1(y_n)} \left[\ln q_1(y_n) - \ln q_0(y_n)\right] + \sum_{n: x_n(m) = -1, q_0(y_n) > q_1(y_n)} \left[\ln q_0(y_n) - \ln q_1(y_n)\right],
\end{eqnarray}
\end{figure*}
and the sum in (\ref{eqn:decoding-rule-form-2}) is
\begin{eqnarray}
    \label{eqn:decoding-rule-form-2-proof}
    \sum_{n: x_n(m) = 1} \ln q_0(y_n) + \sum_{n: x_n(m) = -1} \ln q_1(y_n).
\end{eqnarray}
So it suffices to show that the sum of (\ref{eqn:decoding-rule-form-1-proof}) and (\ref{eqn:decoding-rule-form-2-proof}) is a constant. Indeed, we have (\ref{eqn:decoder-equivalence}),
\begin{figure*}
\begin{eqnarray}\label{eqn:decoder-equivalence}
    &&\sum_{n: x_n(m) = 1, q_0(y_n) < q_1(y_n)} \left[\ln q_1(y_n) - \ln q_0(y_n)\right] + \sum_{n: x_n(m) = -1, q_0(y_n) > q_1(y_n)} \left[\ln q_0(y_n) - \ln q_1(y_n)\right] \nonumber\\
    &&\quad\quad + \sum_{n: x_n(m) = 1} \ln q_0(y_n) + \sum_{n: x_n(m) = -1} \ln q_1(y_n)\nonumber\\
    &=& \sum_{n: x_n(m) = 1, q_0(y_n) < q_1(y_n)} \ln q_1(y_n) + \sum_{n: x_n(m) = 1, q_0(y_n) > q_1(y_n)} \ln q_0(y_n) \nonumber\\
    &&\quad\quad + \sum_{n: x_n(m) = -1, q_0(y_n) > q_1(y_n)} \ln q_0(y_n) + \sum_{n: x_n(m) = -1, q_0(y_n) < q_1(y_n)} \ln q_1(y_n)\nonumber\\
    &=& \sum_{n: q_0(y_n) > q_1(y_n)} \ln q_0(y_n) + \sum_{n: q_0(y_n) < q_1(y_n)} \ln q_1(y_n) = \sum_{n = 1}^N \max\left\{\ln q_0(y_n), \ln q_1(y_n)\right\},
\end{eqnarray}
\end{figure*}
which is a constant independent of the specific codeword. Since (\ref{eqn:decoding-rule-form-2}) is the ML decoder, this proves the optimality of the decoding rule (\ref{eqn:decoding-rule-form-1}). \textbf{Q.E.D.}

Now let us apply Lemma \ref{lem:optimality-decoding-rule} to the case of antipodal input over a real AWGN channel. We map $\sqrt{P}$ to the letter $1$ and $- \sqrt{P}$ to the letter $-1$, so that $q_0(y) = \phi(y - \sqrt{P})$ and $q_1(y) = \phi(y + \sqrt{P})$. Consequently we have $\ln \frac{q_0(y_n)}{q_1(y_n)} = 2 \sqrt{P} y_n$. This establishes the optimality of $r_n(\underline{y}) = |y_n|$ in (\ref{eqn:decoding-rule}).

\subsection{Proof of Lemma \ref{lem:ED1}}
\label{appendix:ED1}

Under $m = 1$, we have
\begin{eqnarray}
    \label{eqn:proof-lem-ED1-1}
    &&\mathbb{E} \rvd(1) \nonumber\\
    &=& \frac{1}{N^2} \sum_{n = 1}^N \mathbb{E} \left[\rvr_n \mathbf{1}(\mathrm{sgn}(\rvx_n(1) + \rvz_n)\cdot \rvx_n(1) < 0)\right].\nonumber\\
\end{eqnarray}
For each summand in (\ref{eqn:proof-lem-ED1-1}), we have
\begin{eqnarray}
    \label{eqn:proof-lem-ED1-2}
    &&\mathbb{E} \left[\rvr_n \mathbf{1}(\mathrm{sgn}(\rvx_n(1) + \rvz_n)\cdot \rvx_n(1) < 0)\right]\nonumber\\
    &=& \frac{1}{2} \mathbb{E} \left[\rvr_n \mathbf{1}(\rvz_n + \sqrt{P} < 0)\Big| \rvx_n(1) \!=\! \sqrt{P}\right] \nonumber\\
    &&+ \frac{1}{2} \mathbb{E} \left[\rvr_n \mathbf{1}(\rvz_n - \sqrt{P} > 0)\Big| \rvx_n(1) \!=\! - \sqrt{P}\right].\nonumber\\
\end{eqnarray}
Taking the first expectation in (\ref{eqn:proof-lem-ED1-2}) and applying the law of total expecation, we have
\begin{eqnarray}
    && \mathbb{E} \left[\rvr_n \mathbf{1}(\rvz_n + \sqrt{P} < 0)\Big| \rvx_n(1) = \sqrt{P}\right] \nonumber\\
    &=& \mathbb{E} \left[\mathbb{E} \left[\rvr_n \mathbf{1}(\rvz_n + \sqrt{P} < 0)\Big| \rvx_n(1) = \sqrt{P}, \rvz_n\right]\right],\nonumber\\
\end{eqnarray}
for which it holds that
\begin{eqnarray}
    \label{eqn:proof-lem-ED1-3}
    &&\mathbb{E} \left[\rvr_n \mathbf{1}(z + \sqrt{P} < 0)\Big| \rvx_n(1) = \sqrt{P}, \rvz_n = z\right]\nonumber\\
    &=& \left\{
        \begin{aligned}
            \mathbb{E} \left[\rvr_n \Big| \rvx_n(1) = \sqrt{P}, \rvz_n = z\right] &\quad& \mbox{if}\; z < - \sqrt{P},\\
            0 &\quad& \mbox{else}.
        \end{aligned}
    \right.\nonumber\\
\end{eqnarray}
Noting that $\rvr_n$ is defined as the rank of $|\rvy_n|$ among the sorted array consisting of $\{|\rvy_1|, |\rvy_2|, \ldots, |\rvy_N|\}$, and that $\{|\rvy_1|, |\rvy_2|, \ldots, |\rvy_N|\}$ are i.i.d. according to Lemma \ref{lem:independent-absY}, we have that the conditional expectation in the first case of (\ref{eqn:proof-lem-ED1-3}) is the expectation of the rank when inserting $- z - \sqrt{P}$ into a sorted array of $N - 1$ i.i.d. samples of $\{|\rvy|\}$. Let us denote this expected rank by $s(z)$, $z < - \sqrt{P}$, and thus
\begin{eqnarray}
    \label{eqn:proof-lem-ED1-4}
    &&\mathbb{E} \left[\rvr_n \mathbf{1}(\rvz_n + \sqrt{P} < 0)\Big| \rvx_n(1) = \sqrt{P}\right] \nonumber\\
    &=& \int_{-\infty}^{-\sqrt{P}} s(z) \phi(z) \mathrm{d}z.
\end{eqnarray}

Similarly, the second expectation in (\ref{eqn:proof-lem-ED1-2}) can also be evaluated in the same way as above, and the result is identical to the right hand side of (\ref{eqn:proof-lem-ED1-4}). So we have
\begin{eqnarray}
    \label{eqn:proof-lem-ED1-5}
    &&\mathbb{E} \left[\rvr_n \mathbf{1}(\mathrm{sgn}(\rvx_n(1) + \rvz_n)\cdot \rvx_n(1) < 0)\right] \nonumber\\
    &=& \int_{-\infty}^{-\sqrt{P}} s(z) \phi(z) \mathrm{d}z,
\end{eqnarray}
which does not depend upon the index $n$. Consequently,
\begin{eqnarray}
    \mathbb{E} \rvd(1) = \int_{-\infty}^{-\sqrt{P}} \frac{s(z)}{N} \phi(z) \mathrm{d}z.
\end{eqnarray}

Applying Lemma \ref{lem:supporting-sz}, we obtain
\begin{eqnarray}
    \lim_{N \rightarrow \infty} \mathbb{E}\rvd(1) &=& \int_{-\infty}^{-\sqrt{P}} \lim_{N \rightarrow \infty} \frac{s(z)}{N} \phi(z) \mathrm{d}z \nonumber\\
    &=& \int_{-\infty}^{-\sqrt{P}} \Psi(-z - \sqrt{P}) \phi(z) \mathrm{d}z,
\end{eqnarray}
which is exactly (\ref{eqn:lem-ED1}) after a change of variable. \textbf{Q.E.D.}

\subsection{Proof of Lemma \ref{lem:varD1}}
\label{appendix:varD1}

Defining $\rvw_n = (\rvr_n/N) \mathbf{1}(\mathrm{sgn}(\rvx_n(1) + \rvz_n) \rvx_n(1) < 0)$ and $\tilde{\rvw}_n = \rvw_n - \mathbb{E} \rvw_n$, we have
\begin{eqnarray}
    \label{eqn:proof-lem-varD1-1}
    \mathrm{var} \rvd(1) = \frac{1}{N^2} \sum_{i = 1}^N \sum_{j = 1}^N \mathbb{E}[\tilde{\rvw}_i \tilde{\rvw}_j].
\end{eqnarray}

For $j = i$, let us calculate $\mathbb{E}[\tilde{\rvw}_i^2] = \mathbb{E}[\rvw_i^2] - \left[\mathbb{E} \rvw_i \right]^2$. In (\ref{eqn:proof-lem-ED1-5}) we have obtained
\begin{eqnarray}
    \label{eqn:proof-lem-varD1-2}
    \mathbb{E} \rvw_i = \int_{-\infty}^{-\sqrt{P}} \frac{s(z)}{N} \phi(z) \mathrm{d}z.
\end{eqnarray}
Following similar steps as those in Appendix \ref{appendix:ED1}, we have
\begin{eqnarray}
    \mathbb{E}[\rvw_i^2] = \int_{-\infty}^{-\sqrt{P}} \frac{s_2(z)}{N^2} \phi(z) \mathrm{d}z,
\end{eqnarray}
where we define $s_2(z)$ as the expectation of the square of the rank when inserting $- z - \sqrt{P}$ into a sorted array of $N - 1$ i.i.d. samples of $\{|\rvy|\}$.

Applying Lemma \ref{lem:supporting-sz}, we then have
\begin{eqnarray}
    \lim_{N \rightarrow \infty} \mathbb{E} \rvw_i &=& \int_{-\infty}^{-\sqrt{P}} \Psi(-z - \sqrt{P}) \phi(z) \mathrm{d}z,\nonumber\\
    \mbox{and}\; \lim_{N \rightarrow \infty} \mathbb{E}[\rvw_i^2] &=& \int_{-\infty}^{-\sqrt{P}} \Psi^2(-z - \sqrt{P}) \phi(z) \mathrm{d}z,\nonumber
\end{eqnarray}
respectively. Hence
\begin{eqnarray}
    \lim_{N \rightarrow \infty} \mathbb{E}[\tilde{\rvw}_i^2] &=& \lim_{N \rightarrow \infty} \mathbb{E}[\rvw_i^2] - \left[\lim_{N \rightarrow \infty}\mathbb{E} \rvw_i \right]^2\nonumber\\
    &=& \int_{-\infty}^{-\sqrt{P}} \Psi^2(-z - \sqrt{P}) \phi(z) \mathrm{d}z \nonumber\\
    &&- \left[\int_{-\infty}^{-\sqrt{P}} \Psi(-z - \sqrt{P}) \phi(z) \mathrm{d}z\right]^2,\nonumber\\
\end{eqnarray}
which is a finite value and is independent of $i$. Since there are $N$ such terms in the sum of (\ref{eqn:proof-lem-varD1-1}), when scaled by $1/N^2$, their total contribution to $\mathrm{var} \rvd(1)$ asymptotically vanishes with $N$ at a rate of $O(1/N)$.

For $j \neq i$, we need to calculate $\mathbb{E}[\tilde{\rvw}_i \tilde{\rvw}_j] = \mathbb{E}[\rvw_i \rvw_j] - \mathbb{E} \rvw_i \mathbb{E} \rvw_j$. We have already derived $\mathbb{E} \rvw_i$ in (\ref{eqn:proof-lem-varD1-2}), which does not depend upon the index $i$. Now let us turn to $\mathbb{E}[\rvw_i \rvw_j]$, for which we have (\ref{eqn:proof-lem-varD1-3}).
\begin{figure*}
\begin{eqnarray}
    \label{eqn:proof-lem-varD1-3}
    \mathbb{E}[\rvw_i \rvw_j] &=& \frac{1}{N^2} \mathbb{E}\left[\rvr_i \rvr_j \mathbf{1}\left(\mathrm{sgn}(\rvx_i(1) + \rvz_i) \rvx_i(1) < 0\right) \mathbf{1}\left(\mathrm{sgn}(\rvx_j(1) + \rvz_j) \rvx_j(1) < 0\right)\right]\nonumber\\
    &=& \frac{1}{4 N^2} \mathbb{E}\left[\rvr_i \rvr_j \mathbf{1}(\rvz_i < -\sqrt{P}) \mathbf{1}(\rvz_j < - \sqrt{P}) \Big| \rvx_i(1) = \sqrt{P}, \rvx_j(1) = \sqrt{P}\right]\nonumber\\
    &+& \frac{1}{4 N^2} \mathbb{E}\left[\rvr_i \rvr_j \mathbf{1}(\rvz_i < -\sqrt{P}) \mathbf{1}(\rvz_j > \sqrt{P}) \Big| \rvx_i(1) = \sqrt{P}, \rvx_j(1) = -\sqrt{P}\right]\nonumber\\
    &+& \frac{1}{4 N^2} \mathbb{E}\left[\rvr_i \rvr_j \mathbf{1}(\rvz_i > \sqrt{P}) \mathbf{1}(\rvz_j < - \sqrt{P}) \Big| \rvx_i(1) = - \sqrt{P}, \rvx_j(1) = \sqrt{P}\right]\nonumber\\
    &+& \frac{1}{4 N^2} \mathbb{E}\left[\rvr_i \rvr_j \mathbf{1}(\rvz_i > \sqrt{P}) \mathbf{1}(\rvz_j > \sqrt{P}) \Big| \rvx_i(1) = - \sqrt{P}, \rvx_j(1) = - \sqrt{P}\right].
\end{eqnarray}
\end{figure*}
By symmetry, it suffices to examine the first summand in (\ref{eqn:proof-lem-varD1-3}) and the values of the other three summands are identical to it. Applying the law of total expectation, we have (\ref{eqn:Rij}),
\begin{figure*}
\begin{eqnarray}\label{eqn:Rij}
    &&\mathbb{E}\left[\rvr_i \rvr_j \mathbf{1}(\rvz_i < -\sqrt{P}) \mathbf{1}(\rvz_j < - \sqrt{P}) \Big| \rvx_i(1) = \sqrt{P}, \rvx_j(1) = \sqrt{P}\right]\nonumber\\
    &=& \mathbb{E}\left[\mathbb{E}\left[\rvr_i \rvr_j \mathbf{1}(\rvz_i < -\sqrt{P}) \mathbf{1}(\rvz_j < - \sqrt{P}) \Big| \rvx_i(1) = \sqrt{P}, \rvx_j(1) = \sqrt{P}, \rvz_i, \rvz_j\right]\right],
\end{eqnarray}
\end{figure*}
wherein the inner conditional expectation, when $\rvz_i = z_i < - \sqrt{P}$ and $\rvz_j = z_j < - \sqrt{P}$, is the expectation of the product of the ranks when inserting $- z_i - \sqrt{P}$ and $- z_j - \sqrt{P}$ into a sorted array of $N - 2$ i.i.d. samples of $\{|\rvy|\}$. Denoting this inner conditional expectation as $\tilde{s}(z_i, z_j)$, we can hence write $\mathbb{E}[\rvw_i \rvw_j]$ as an integral
\begin{eqnarray}
    \mathbb{E}[\rvw_i \rvw_j] = \int_{-\infty}^{-\sqrt{P}} \int_{-\infty}^{-\sqrt{P}} \frac{\tilde{s}(z_i, z_j)}{N^2} \phi(z_i) \phi(z_j) \mathrm{d}z_i \mathrm{d}z_j.
\end{eqnarray}
Applying Lemma \ref{lem:supporting-stilde}, we can obtain (\ref{eqn:EWiWj}).
\begin{figure*}
\begin{eqnarray}\label{eqn:EWiWj}
    \mathbb{E}[\rvw_i \rvw_j] &=& \int_{-\infty}^{-\sqrt{P}} \int_{-\infty}^{-\sqrt{P}} \Psi(- z_i - \sqrt{P}) \Psi(- z_j - \sqrt{P}) \phi(z_i) \phi(z_j) \mathrm{d}z_i \mathrm{d}z_j + O\left(\frac{1}{N}\right)\nonumber\\
    &=& \int_{-\infty}^{-\sqrt{P}} \Psi(- z_i - \sqrt{P}) \phi(z_i) \mathrm{d}z_i \int_{-\infty}^{-\sqrt{P}} \Psi(- z_j - \sqrt{P}) \phi(z_j) \mathrm{d}z_j + O\left(\frac{1}{N}\right)\nonumber\\
    &=& \mathbb{E} \rvw_i \mathbb{E} \rvw_j + O\left(\frac{1}{N}\right).
\end{eqnarray}
\end{figure*}
Therefore, $\mathbb{E}[\tilde{\rvw}_i \tilde{\rvw}_j] = \mathbb{E}[\rvw_i \rvw_j] - \mathbb{E} \rvw_i \mathbb{E} \rvw_j = O(1/N)$ for any $i \neq j$. There are $N(N - 1)$ such terms in the sum of (\ref{eqn:proof-lem-varD1-1}). When scaled by $1/N^2$, their total contribution to $\mathrm{var} \rvd(1)$ asymptotically vanishes with $N$ at a rate of $O(1/N)$.

So in summary, as $N \rightarrow \infty$, $\mathrm{var} \rvd(1)$ asymptotically tends towards zero. \textbf{Q.E.D.}

\subsection{Proof of Lemma \ref{lem:EDm}}
\label{appendix:EDm}

For any $m' \neq 1$, we have
\begin{eqnarray}
    \label{eqn:proof-lem-EDm-1}
    &&\mathbb{E}\left\{e^{N \theta \rvd(m')} \Big| \underline{\rvy}\right\}\nonumber\\
    &=& \mathbb{E}\left\{e^{\theta \sum_{n = 1}^N (\rvr_n/N) \mathbf{1}(\mathrm{sgn}(\rvy_n) \rvx_n(m') < 0)} \Big| \underline{\rvy}\right\}\nonumber\\
    &=& \mathbb{E}\left\{\prod_{n = 1}^N e^{\theta (\rvr_n/N) \mathbf{1}(\mathrm{sgn}(\rvy_n) \rvx_n(m') < 0)} \Big| \underline{\rvy}\right\},
\end{eqnarray}
wherein we emphasize that $\underline{\rvy}$ is induced by $\underline{\rvx}(1)$, which is hence independent of $\underline{\rvx}(m')$. So we can continue the evaluation of (\ref{eqn:proof-lem-EDm-1}) as 
\begin{eqnarray}
    \label{eqn:proof-lem-EDm-2}
    &&\mathbb{E}\left\{e^{N \theta \rvd(m')} \Big| \underline{\rvy}\right\} \nonumber\\
    &=& \prod_{n = 1}^N \mathbb{E}\left\{e^{\theta (\rvr_n/N) \mathbf{1}(\mathrm{sgn}(\rvy_n) \rvx_n(m') < 0)} \Big| \underline{\rvy}\right\}.
\end{eqnarray}

For each term in (\ref{eqn:proof-lem-EDm-2}), we have
\begin{eqnarray}
    &&\mathbb{E}\left\{e^{\theta (\rvr_n/N) \mathbf{1}(\mathrm{sgn}(\rvy_n) \rvx_n(m') < 0)} \Big| \underline{\rvy}\right\} \nonumber\\
    &=& \frac{1}{2} e^{\theta (\rvr_n/N) \mathbf{1}(\rvy_n < 0)} + \frac{1}{2} e^{\theta (\rvr_n/N) \mathbf{1}(\rvy_n > 0)}\nonumber\\
    &=& \frac{1}{2} \left(1 + e^{\theta \rvr_n/N}\right),
\end{eqnarray}
where we utilize the fact that conditioned upon $\underline{\rvy}$, $\rvr_n$ is determinisitc and hence we may remove the conditional expectation operator.

Returning to (\ref{eqn:proof-lem-EDm-2}) and taking its logarithm, we obtain
\begin{eqnarray}
    &&\frac{1}{N} \ln \mathbb{E}\left\{e^{N \theta \rvd(m')} \Big| \underline{\rvy}\right\} \nonumber\\
    &=& \frac{1}{N} \sum_{n = 1}^N \ln \left(1 + e^{\theta \rvr_n/N}\right) - \ln 2\nonumber\\
    &=& \frac{1}{N} \sum_{n = 0}^{N - 1} \ln \left(1 + e^{\theta n/N}\right) - \ln 2,
\end{eqnarray}
where we utilize the fact that $\{\rvr_n\}_{n = 1, \ldots, N}$ is simply a permutation of $\{0, 1, \ldots, N - 1\}$. Passing to the limit of $N \rightarrow \infty$, the series sum becomes an integration, and we have
\begin{eqnarray}
    \Lambda(\theta) &=& \lim_{N \rightarrow \infty} \frac{1}{N} \ln \mathbb{E}\left\{e^{N \theta \rvd(m')} \Big| \underline{\rvy}\right\}\nonumber\\
    &=& \lim_{N \rightarrow \infty} \frac{1}{N} \sum_{n = 0}^{N - 1} \ln \left(1 + e^{\theta n/N}\right) - \ln 2\nonumber\\
    &=& \int_0^1 \ln \left(1 + e^{\theta t}\right) \mathrm{d}t - \ln 2.
\end{eqnarray}
This completes the proof of Lemma \ref{lem:EDm}. \textbf{Q.E.D.}

\subsection{GMI of cdf-GRAND}
\label{appendix:cdf-GRAND}

For the decoding rule of cdf-GRAND, we can readily apply the formula of GMI \cite[Eqn. (12)]{ganti00:it} as (\ref{eqn:cdf-GRAND}),
\begin{figure*}
\begin{eqnarray}
    \label{eqn:cdf-GRAND}
    I_\mathrm{cdf-GRAND} = \sup_{\theta < 0} \left\{\theta \mathbb{E} d(\rvx, \rvy) - \mathbb{E}\left[\ln \sum_{x \in \{\sqrt{P}, - \sqrt{P}\}} \frac{e^{\theta d(x, \rvy)}}{2} \right]\right\},
\end{eqnarray}
\end{figure*}
where $d(x, y) = \Psi(|y|) \cdot \mathbf{1}(\mathrm{sgn}(y)\cdot x < 0)$, and the expectations are with respect to a joint probability distribution induced according to the AWGN channel law (\ref{eqn:AWGN-channel}).

We have
\begin{eqnarray}
    \label{eqn:cdf-GRAND-1}
    &&\mathbb{E} d(\rvx, \rvy) \nonumber\\
    &=& \frac{1}{2} \mathbb{E} d(- \sqrt{P}, - \sqrt{P} + \rvz) + \frac{1}{2} \mathbb{E} d(\sqrt{P}, \sqrt{P} + \rvz)\nonumber\\
    &=& \frac{1}{2} \mathbb{E} \left[\Psi(|\rvz - \sqrt{P}|) \mathbf{1}(\rvz > \sqrt{P})\right] \nonumber\\
    && \quad + \frac{1}{2} \mathbb{E} \left[\Psi(|\rvz + \sqrt{P}|) \mathbf{1}(\rvz < - \sqrt{P})\right]\nonumber\\
    &=& \frac{1}{2} \int_{\sqrt{P}}^\infty \Psi(z - \sqrt{P}) \phi(z) \mathrm{d}z \nonumber\\
    && \quad + \frac{1}{2} \int_{-\infty}^{-\sqrt{P}} \Psi(- z - \sqrt{P}) \phi(z) \mathrm{d}z\nonumber\\
    &=& \int_0^\infty \Psi(t) \phi(t + \sqrt{P}) \mathrm{d}t.
\end{eqnarray}

Next, since it holds that
\begin{eqnarray}
    &&\sum_{x \in \{\sqrt{P}, - \sqrt{P}\}} \frac{e^{\theta d(x, \rvy)}}{2} \nonumber\\
    &=& \frac{1}{2} \left(e^{\theta \Psi(|\rvy|) \mathbf{1}(\rvy < 0)} + e^{\theta \Psi(|\rvy|) \mathbf{1}(\rvy > 0)}\right)\nonumber\\
    &=& \frac{1}{2} \left(e^{\theta \Psi(|\rvy|)} + 1\right),
\end{eqnarray}
we have
\begin{eqnarray}
    \label{eqn:cdf-GRAND-2}
    &&\mathbb{E}\left[\ln \sum_{x \in \{\sqrt{P}, - \sqrt{P}\}} \frac{e^{\theta d(x, \rvy)}}{2} \right] \nonumber\\
    &=& \mathbb{E}\left[\ln \left(e^{\theta \Psi(|\rvy|)} + 1\right)\right] - \ln 2\nonumber\\
    &=& \int_0^1 \ln (1 + e^{\theta t}) \mathrm{d}t - \ln 2,
\end{eqnarray}
where the last equality follows from the fact that the cdf random variable $\Psi(|\rvy|)$ follows a uniform distribution over the unit interval.

Putting (\ref{eqn:cdf-GRAND-1}) and (\ref{eqn:cdf-GRAND-2}) back to (\ref{eqn:cdf-GRAND}), we obtain the relationship $I_\mathrm{cdf-GRAND} = I_\mathrm{ORBGRAND}$. \textbf{Q.E.D.}

\subsection{Supporting Lemmas}
\label{appendix:supporting}

\begin{lem}
    \label{lem:supporting-sz}
    Consider inserting a real number $v$ into a sorted array consisting of $N - 1$ i.i.d. real random variables $\{\rvt_i\}_{i = 1, \ldots, N - 1}$ with cdf $F(t)$. Then the expectation of the rank of $v$, $s(v)$, satisfies
    \begin{eqnarray}
        \label{eqn:supporting-sz}
        \lim_{N \rightarrow \infty} \frac{s(v)}{N} = F(v),
    \end{eqnarray}
    and the expectation of the square of the rank of $v$, $s_2(v)$, satisfies
    \begin{eqnarray}
        \label{eqn:supporting-s2z}
        \lim_{N \rightarrow \infty} \frac{s_2(v)}{N^2} = F^2(v),
    \end{eqnarray}
\end{lem}
\textit{Proof:} Denote the rank of $v$ after insertion as $\rvr$. It is clear that $\rvr - 1$ obeys a binomial distribution with parameters $N - 1$ and $F(v)$. So the expectation $s(v)$ is $(N - 1) F(v) + 1$, and consequently (\ref{eqn:supporting-sz}) holds. In order to prove (\ref{eqn:supporting-s2z}), we note that the second order moment of this binomial distribution is given by $(N - 1)^2 F^2(v) + (N - 1) F(v) [1 - F(v)]$. So as $N \rightarrow \infty$, all the lower order terms in $s_2(v)/N^2$ asymptotically vanish, leaving $F^2(v)$ only. \textbf{Q.E.D.}

\begin{lem}
    \label{lem:supporting-stilde}
    Consider inserting two real numbers $v_a$ and $v_b$ into a sorted array consisting of $N - 2$ i.i.d. real random variables $\{\rvt_i\}_{i = 1, \ldots, N - 2}$ with cdf $F(t)$. Then the expectation of the product of the ranks of $v_a$ and $v_b$, $\tilde{s}(v_a, v_b)$, satisfies
    \begin{eqnarray}
        \label{eqn:supporting-stilde}
        \frac{\tilde{s}(v_a, v_b)}{N^2} = F(v_a) F(v_b) + O\left(\frac{1}{N}\right).
    \end{eqnarray}
\end{lem}
\textit{Proof:} Without loss of generality, assume that $v_a < v_b$. Denote the ranks of $v_a$ and $v_b$ when inserted into the sorted array as $\rvr_a$ and $\rvr_b$, respectively. Clearly, the probability of $(\rvr_a = r_a, \rvr_b = r_b)$ is
\begin{eqnarray}
    &&\frac{(N - 2)!}{(r_a - 1)! \cdot (r_b - r_a - 1)! \cdot (N - r_b)!} \nonumber\\
    && \times \left[F(v_a)\right]^{r_a - 1} \left[F(v_b) - F(v_a)\right]^{r_b - r_a - 1} \left[1 - F(v_b)\right]^{N - r_b};\nonumber\\
\end{eqnarray}
that is, $(\rvr_a - 1, \rvr_b - \rvr_a - 1, N - \rvr_b)$ obeying a multinomial distribution. We hence have (\ref{eqn:tilde-s-eval}),
\begin{figure*}
\begin{eqnarray}\label{eqn:tilde-s-eval}
    \tilde{s}(v_a, v_b) &=& \mathbb{E}\left[\rvr_a \rvr_b\right]\nonumber\\
    &=& \mathbb{E}\left[(\rvr_a - 1)(\rvr_b - \rvr_a - 1)\right] + \mathbb{E}\left[(\rvr_a - 1)^2\right] + 3\mathbb{E}[\rvr_a - 1] + \mathbb{E}[\rvr_b - \rvr_a - 1] + 2\nonumber\\
    &=& (N - 2)^2 F(v_a) [F(v_b) - F(v_a)] - (N - 2) F(v_a) [F(v_b) - F(v_a)] + (N - 2)^2 F(v_a)^2 \nonumber\\
    &&\quad\quad + (N - 2) F(v_a) [1 - F(v_a)] + 3 (N - 2) F(v_a) + (N - 2) [F(v_b) - F(v_a)] + 2\nonumber\\
    &=& (N - 2)^2 F(v_a) F(v_b) - (N - 2) [F(v_a) F(v_b) - 3 F(v_a) - F(v_b)] + 2,
\end{eqnarray}
\end{figure*}
where we utilize Lemma \ref{lem:supporting-sz} and the covariance of multinomial distribution. Consequently we obtain (\ref{eqn:supporting-stilde}) by passing to the limit of $N \rightarrow \infty$. \textbf{Q.E.D.}

\end{document}